\documentclass[conference]{IEEEtran}
\IEEEoverridecommandlockouts
\usepackage{cite}
\usepackage{amsmath,amssymb,amsfonts}
\usepackage{algorithmic}
\usepackage{graphicx}
\usepackage{textcomp}
\usepackage{xcolor}
\usepackage{acronym}
\usepackage{bm}
\usepackage{mleftright}
\usepackage{tikz}
\usepackage{pgfplots}
\usepackage{booktabs}
\pgfplotsset{compat=1.18}
\usetikzlibrary{arrows, arrows.meta, positioning, quotes, shapes, fit, calc, decorations.pathreplacing, calligraphy, plotmarks, external, pgfplots.groupplots, spy}
\tikzstyle{block} = [draw, thick, text=black, align=center, minimum width=1.2cm, minimum height=0.6cm, font=\small]
\tikzstyle{Block} = [draw, thick, rounded corners, text=black, align=center, minimum width=2cm, minimum height=1.2cm, font=\small]
\usepackage{balance}
\usepackage{siunitx}
\DeclareSIUnit\operations{Op}
\DeclareSIUnit\BL{\giga\bit\kilo\meter\per\second}
\DeclareSIUnit\FL{\decibel\per\kilo\meter}
\DeclareSIUnit\CD{\pico\second\per\nano\meter\per\kilo\meter}
\DeclareSIUnit\baud{Bd}
\DeclareSIUnit{\nothing}{\relax}

\definecolor{RPTU_BlueGray}{RGB}{80,114,137}
\definecolor{RPTU_GreenGray}{RGB}{119,182,186}
\definecolor{RPTU_DarkBlue}{RGB}{4,44,88}
\definecolor{RPTU_LightBlue}{RGB}{106,178,231}
\definecolor{RPTU_DarkGreen}{RGB}{0,107,107}
\definecolor{RPTU_LightGreen}{RGB}{38,208,124}
\definecolor{RPTU_Violett}{RGB}{76,53,117}
\definecolor{RPTU_Pink}{RGB}{209,56,150}
\definecolor{RPTU_Red}{RGB}{227,27,76}
\definecolor{RPTU_Orange}{RGB}{255,162,82}
\definecolor{RPTU_Black}{RGB}{0,0,0}
\definecolor{RPTU_White}{RGB}{255,255,255}
\definecolor{Navy_Blue}{RGB}{0,0,128}
\definecolor{yellow}{RGB}{204,204,0}
\definecolor{green}{RGB}{0,204,0}

\pgfdeclarelayer{background}
\pgfdeclarelayer{background_1}
\pgfdeclarelayer{background_2}
\pgfdeclarelayer{background_3}
\pgfdeclarelayer{foreground_6}
\pgfdeclarelayer{foreground_5}
\pgfdeclarelayer{foreground_4}
\pgfdeclarelayer{foreground_3}
\pgfdeclarelayer{foreground_2}
\pgfdeclarelayer{foreground_1}
\pgfdeclarelayer{foreground}
\pgfsetlayers{background,background_1,background_2,background_3,main,foreground_6,foreground_5,foreground_4,foreground_3,foreground_2,foreground_1,foreground}

\DeclareMathSymbol{\mlq}{\mathord}{operators}{``}
\DeclareMathSymbol{\mrq}{\mathord}{operators}{`'}

\def\thus{\relax
	\ifmmode
	\implies
	\else
	$\implies$
	\fi}

\def\endofproof{
	\ifmmode
		\text{\hfill} \textcolor{KITgreen}{\rule{2mm}{2mm}}
	\else
		$\hfill \textcolor{KITgreen}{\rule{2mm}{2mm}}$
	\fi}

\def\rz{\ifmmode{\mathbb{R}}%
    \else{\hbox{$\mathbb{R}$}}\fi} 
\def\nz{\ifmmode{\mathbb{N}}%
    \else{\hbox{$\mathbb{N}$}}\fi} 
\def\gz{\ifmmode{\mathbb{Z}}%
   \else{\hbox{$\mathbb{Z}$}}\fi} 
\def\cz{\ifmmode{\mathbb{C}}
    \else{\hbox{$\mathbb{C}$}}\fi}%
\def\qz{\ifmmode{\mathbb{Q}}%
    \else{\hbox{$\mathbb{Q}$}}\fi}%
\def\K{\ifmmode{\mathbb{K}}%
    \else{\hbox{$\mathbb{K}$}}\fi}%

\def\Er{\ifmmode{\mathbb{E}}%
    \else{\hbox{$\mathbb{E}$}}\fi}%
    
\def\V{
	\ifmmode
	{\mathrm{V}}
	\else
	${\mathrm{V}}$
	\fi}

\usepackage{trfsigns}

\makeatletter
\newcommand*{\AdjustMargins}{%
    \setlength{\beamer@rightmargin}{0em}%
    \setlength{\beamer@leftmargin}{0em}%
}
\makeatother

\newcommand{\hi}[2][KITyellow]{
	\ifmmode
	{
		 \mathchoice%
			{\colorbox{#1!50}{\textcolor{black}{$\displaystyle#2$}}}%
			{\colorbox{#1!50}{\textcolor{black}{$\textstyle#2$}}}
			{\colorbox{#1!50}{\textcolor{black}{$\scriptstyle#2$}}}
			{\colorbox{#1!50}{\textcolor{black}{$\scriptscriptstyle#2$}}}
	}
	\else
	{
		{\colorbox{#1!50}{\textcolor{black}{#2}}}
	}
	\fi
	}

\newcommand{\footnotetextoff}[2]{\alt<#1>{\let\thefootnote\relax\footnotetext{~}}{\footnotetext{\scriptsize #2}}}

\definecolor{kit-green100}{rgb}{0,.59,.51}
\definecolor{kit-green70}{rgb}{.3,.71,.65}
\definecolor{kit-green50}{rgb}{.50,.79,.75}
\definecolor{kit-green30}{rgb}{.69,.87,.85}
\definecolor{kit-green15}{rgb}{.85,.93,.93}
\definecolor{KITgreen}{rgb}{0,.59,.51}

\definecolor{KITpalegreen}{RGB}{130,190,60}
\colorlet{kit-maigreen100}{KITpalegreen}
\colorlet{kit-maigreen70}{KITpalegreen!70}
\colorlet{kit-maigreen50}{KITpalegreen!50}
\colorlet{kit-maigreen30}{KITpalegreen!30}
\colorlet{kit-maigreen15}{KITpalegreen!15}

\definecolor{KITblue}{rgb}{.27,.39,.66}
\definecolor{kit-blue100}{rgb}{.27,.39,.67}
\definecolor{kit-blue70}{rgb}{.49,.57,.76}
\definecolor{kit-blue50}{rgb}{.64,.69,.83}
\definecolor{kit-blue30}{rgb}{.78,.82,.9}
\definecolor{kit-blue15}{rgb}{.89,.91,.95}

\definecolor{KITyellow}{rgb}{.98,.89,0}
\definecolor{kit-yellow100}{cmyk}{0,.05,1,0}
\definecolor{kit-yellow70}{cmyk}{0,.035,.7,0}
\definecolor{kit-yellow50}{cmyk}{0,.025,.5,0}
\definecolor{kit-yellow30}{cmyk}{0,.015,.3,0}
\definecolor{kit-yellow15}{cmyk}{0,.0075,.15,0}

\definecolor{KITorange}{rgb}{.87,.60,.10}
\definecolor{kit-orange100}{cmyk}{0,.45,1,0}
\definecolor{kit-orange70}{cmyk}{0,.315,.7,0}
\definecolor{kit-orange50}{cmyk}{0,.225,.5,0}
\definecolor{kit-orange30}{cmyk}{0,.135,.3,0}
\definecolor{kit-orange15}{cmyk}{0,.0675,.15,0}

\definecolor{KITred}{rgb}{.63,.13,.13}
\definecolor{kit-red100}{cmyk}{.25,1,1,0}
\definecolor{kit-red70}{cmyk}{.175,.7,.7,0}
\definecolor{kit-red50}{cmyk}{.125,.5,.5,0}
\definecolor{kit-red30}{cmyk}{.075,.3,.3,0}
\definecolor{kit-red15}{cmyk}{.0375,.15,.15,0}

\definecolor{KITlilac}{RGB}{160,0,120}
\colorlet{kit-lila100}{KITlilac}
\colorlet{kit-lila70}{KITlilac!70}
\colorlet{kit-lila50}{KITlilac!50}
\colorlet{kit-lila30}{KITlilac!30}
\colorlet{kit-lila15}{KITlilac!15}

\definecolor{KITcyanblue}{RGB}{80,170,230}
\colorlet{kit-cyanblue100}{KITcyanblue}
\colorlet{kit-cyanblue70}{KITcyanblue!70}
\colorlet{kit-cyanblue50}{KITcyanblue!50}
\colorlet{kit-cyanblue30}{KITcyanblue!30}
\colorlet{kit-cyanblue15}{KITcyanblue!15}

\definecolor{ALUColor4}{RGB}{255,127,0}   
\definecolor{ALUColor3}{RGB}{77,175,74}   
\definecolor{ALUColor7}{RGB}{153,153,153}   
\definecolor{ALUColor2}{RGB}{55,126,184}   
\definecolor{ALUColor5}{RGB}{152,78,163}   
\definecolor{ALUColor6}{RGB}{166,86,40}    
\definecolor{ALUColor1}{RGB}{228,26,28}   

\definecolor{ALUColor1v}{RGB}{253,204,138}   
\definecolor{ALUColor2v}{RGB}{252,141,89}   
\definecolor{ALUColor3v}{RGB}{215,48,31}   

\def\BibTeX{{\rm B\kern-.05em{\sc i\kern-.025em b}\kern-.08em
    T\kern-.1667em\lower.7ex\hbox{E}\kern-.125emX}}
\begin{document}

\title{Efficient FPGA Implementation of an Optimized SNN-based DFE for Optical Communications\\

\thanks{This work was funded by the German Federal Ministry of Education and Research (BMBF) under grant agreements 16KIS1316 (AI-NET-ANTILLAS) and 16KISK004 (Open6GHuB). Further it was funded by the Carl Zeiss Stiftung under the Sustainable Embedded AI project (P2021-02-009).}
}

\author{
\IEEEauthorblockN{Mohamed Moursi, Jonas Ney, Bilal Hammoud and Norbert Wehn}
\IEEEauthorblockA{\textit{Microelectronic Systems Design (EMS) RPTU, Kaiserslautern-Landau}, Germany\\ Email:\{mmoursi, jonas.ney, bilal.hammoud, norbert.wehn\}@rptu.de}
}

\maketitle
\acrodef{ai}[AI]{artificial intelligence}
\acrodef{ask}[ASK]{amplitude-shift keying}
\acrodef{awgn}[AWGN]{additive white Gaussian noise}
\acrodef{ann}[ANN]{artificial neural network}
\acrodef{asic}[ASIC]{application-specific integrated circuit}
\acrodef{qat}[QAT]{Quantization Aware Training}
\acrodef{ptq}[PTQ]{Post-Training Quantization}
\acrodef{pl}[PL]{Programmable Logic}
\acrodef{ps}[PS]{Programmable System}

\acrodef{bce}[BCE]{binary cross-entropy}
\acrodef{ber}[BER]{bit error rate}
\acrodef{bler}[BLER]{block error rate}
\acrodef{bpsk}[BPSK]{binary phase-shift keying}
\acrodef{bram}[BRAM]{block random access memory}
\acrodef{bilstm}[biLSTM]{bidirectional long short-term memory}
\acrodef{bp}[BP]{backward pass}

\acrodef{cd}[CD]{chromatic dispersion}
\acrodef{cir}[CIR]{combined impulse response}
\acrodef{cma}[CMA]{constant modulus algorithm}
\acrodef{cnn}[CNN]{convolutional neural network}
\acrodef{cpu}[CPU]{central processing unit}
\acrodef{ctp}[CTP]{channel transition probability}
\acrodefplural{ctp}[CTPs]{channel transition probabilities}
\acrodef{csi}[CSI]{channel state information}
\acrodef{conv1d}[conv1d]{one-dimensional convolution}

\acrodef{dnn}[DNN]{deep neural network}
\acrodef{dsp}[DSP]{digital signal processor}
\acrodef{dram}[DRAM]{dynamic random access memory}
\acrodef{dfe}[DFE]{decision-feedback equalizer}
\acrodef{dop}[DOP]{degree of paralellism}
\acrodef{dbp}[DBP]{digital back-propagation}

\acrodef{fc}[FC]{fully connected}
\acrodef{fec}[FEC]{forward error correction}
\acrodef{fir}[FIR]{finite impulse response}
\acrodef{fifo}[FIFO]{first in first out}
\acrodef{fpga}[FPGA]{field programmable gate array}
\acrodef{fp}[FP]{forward pass}
\acrodef{ff}[FF]{flip-flop}

\acrodef{elu}[ELU]{exponential linear unit}

\acrodef{gan}[GAN]{generative adversarial network}
\acrodef{gpu}[GPU]{graphics processing unit}
\acrodef{gops}[GOPS]{giga operations per second}

\acrodef{hls}[HLS]{high-level synthesis}
\acrodef{hwa}[HWA]{historical weight averaging}

\acrodef{imdd}[IM/DD]{intensity-modulation direct detection}
\acrodef{iot}[IoT]{Internet of things}
\acrodef{isi}[ISI]{inter-symbol interference}

\acrodef{ldpc}[LDPC]{low-density parity-check}
\acrodef{lms}[LMS]{least-mean-square}
\acrodef{lut}[LUT]{look-up table}
\acrodef{lif}[LIF]{Leaky-Integrate-and-Fire}

\acrodef{mlsd}[MLSD]{maximum-likelihood sequence detection}
\acrodef{mse}[MSE]{mean-squared-error}
\acrodef{mlse}[MLSE]{maximum likelihood sequence estimator}
\acrodef{mac}[MAC]{multiply-accumulate}
\acrodef{ma}[MA]{moving average}
\acrodef{map}[MAP]{maximum a posteriori}
\acrodef{msm}[MSM]{merge stream module}
\acrodef{mzm}[MZM]{mach-zehnder modulator}
\acrodef{nn}[NN]{neural network}
\acrodef{ns}[NS]{non-saturating}

\acrodef{ogm}[OGM]{overlap generate module}
\acrodef{orm}[ORM]{overlap remove module}
\acrodef{onu}[ONU]{optical network unit}
\acrodef{olt}[OLT]{optical line terminal}
\acrodef{odn}[ODN]{optical distribution network}

\acrodef{pam}[PAM]{pulse-amplitude modulation}
\acrodef{pdf}[pdf]{probability density function}
\acrodef{pe}[PE]{processing element}
\acrodef{pmf}[pmf]{probability mass function}
\acrodef{pon}[PON]{passive optical network}

\acrodef{qlf}[QLF]{quantization loss factor}
\acrodef{qam}[QAM]{quadrature amplitude modulation}

\acrodef{rc}[RC]{raised-cosine}
\acrodef{rrc}[RRC]{root-raised-cosine}
\acrodef{relu}[ReLU]{rectified linear unit}
\acrodef{rls}[RLS]{recursive least squares}
\acrodef{rnn}[RNN]{recurrent neural network}

\acrodef{ser}[SER]{symbol error rate}
\acrodef{sgd}[SGD]{stochastic gradient descent}
\acrodef{sld}[SLD]{square-law detection}
\acrodef{snr}[SNR]{signal-to-noise ratio}
\acrodef{sps}[sps]{samples per symbol}
\acrodef{ssmf}[SSMF]{standard single-mode fiber}
\acrodef{simd}[SIMD]{single instruction multiple data}
\acrodef{ssm}[SSM]{split stream module}
\acrodef{spb}[SPB]{symbols per batch}
\acrodef{slda}[SLDA]{streaming linear discriminant analysis}
\acrodef{snn}[SNN]{spiking neural network}

\acrodef{ti}[TI]{training iteration}

\acrodef{vnle}[VNLE]{Volterra-based nonlinear equalizer}
\begin{abstract}
The ever-increasing demand for higher data rates in communication systems intensifies the need for advanced non-linear equalizers capable of higher performance. Recently \acp{ann} were introduced as a viable candidate for advanced non-linear equalizers, as they outperform traditional methods. However, they are computationally complex and therefore power hungry. \Acp{snn} started to gain attention as an energy-efficient alternative to \acp{ann}. Recent works proved that they can outperform \acp{ann} at this task. In this work, we explore the design space of an \ac{snn}-based \ac{dfe} to reduce its computational complexity for an efficient implementation on \ac{fpga}. Our Results prove that it achieves higher communication performance than \ac{ann}-based \ac{dfe} at roughly the same throughput and at $25\times$ higher energy efficiency.
\end{abstract}


\section{Introduction}
Data rates of modern communication systems increased dramatically in recent years. This leads to severe degradations of the communication performance caused by noise, \ac{isi}, and hardware impairments, increasing the demand for advanced equalizers capable of migrating non-linear distortions. 

Latest research showed that \acp{ann} are a promising candidate to solve the problem of channel equalization for next-gen communication systems~\cite{comparativeStudy, lstm_ch_eq, elm_ch_eq, DL_OFDM_ch_eq}. Since \acp{ann} are able to compensate for non-linearities, they can achieve higher performance than equalizers based on traditional linear filters. Additionally, they are highly flexible and can be designed to adapt to channel variations by retraining~\cite{ney_learn_1, ney_learn_2}. However, compared to traditional methods, \acp{ann} are computationally complex, which makes efficient hardware implementation challenging. Especially on embedded devices, the practical application of \acp{ann} is limited due to the high power consumption and energy requirements.

To solve the problem of stringent power and energy requirements of \acp{ann}, researchers started investigating the use of \acp{snn}, which are inspired by the human brain to provide high energy efficiency. Two characteristics help to achieve this goal. Firstly, neurons in an \ac{snn} communicate using binary spikes instead of high-precision activations commonly used in \acp{ann}. Secondly, \acp{snn} are event-based networks. Hence, neurons only process information if they receive an input. Recent works showed that \acp{snn} can even outperform \acp{ann} for the task of equalization of an optical communication channel~\cite{vonbank2023spiking, arnold2022, bansbach2023spiking}. 

A commonly used platform for the implementation of digital signal processing algorithms are \acp{fpga} since they enable parallel processing, allow for arbitrary precision and are highly flexible. In contrast to neuromorphic hardware, often used for the implementation of \acp{snn}, \acp{fpga} are widely used in real-world communication systems and are therefore a feasible solution for the implementation of the \ac{snn}-based equalizer. 

In this work, we take a first step towards the practical application of \ac{snn}-based equalization by providing an efficient \ac{fpga} implementation of the equalizer investigated on an algorithmic level in \cite{vonbank2023spiking}. A detailed \ac{snn} design-space-exploration is performed and a novel hardware implementation is presented. 

The new contributions of our work are: \begin{itemize}
    \item We perform a detailed design-space-exploration of the \ac{snn} topology and a trade-off analysis of complexity vs. communication performance.
    \item We present a custom, flexible \ac{fpga} architecture of the \ac{lif} neuron, commonly used in \acp{snn}. 
    \item We evaluate the hardware performance of the \ac{snn}-based equalizer and compare it to its \ac{ann} counterpart regarding resource utilization, power consumption and energy efficiency.
    \item We show that \ac{snn} implementations can outperform \ac{ann} implementations in terms of energy efficiency even on synchronous, digital platforms like \acp{fpga} while maintaining better communication performance. Thus, our \ac{snn} implementation outperforms state-of-the-art regarding communication performance and hardware energy efficiency. 
    \item To the best of our knowledge this is the first \ac{fpga} implementation of an \ac{snn}-based \ac{dfe} for optical communications.
\end{itemize}
\section{Channel Model}

\begin{figure}
\begin{center}
	\tikzsetnextfilename{ChannelModel}

\begin{tikzpicture}[node distance=0.2,>=latex, scale=0.7]
    \pgfdeclarelayer{background}
    \pgfdeclarelayer{foreground}
    \pgfsetlayers{main,foreground}

    \def\minWid{0.6cm}; \def\minHei{0.5cm}; \def\arrowLen{0.5cm};
    \def\boxFontSize{\footnotesize}
    
    \begin{pgfonlayer}{foreground}
    \node[block, rounded corners, minimum width=\minWid, minimum height=\minHei] (transmitter) {T};

	\node[block, rounded corners, right=\arrowLen of transmitter, minimum width=\minWid] (txfilter) {RRC};
	\node[block, rounded corners, right=\arrowLen of txfilter, minimum width=\minWid] (channel) {CD};
	\node[block, rounded corners, right=\arrowLen of channel, minimum width=\minWid] (sld) {$|\cdot|^2$};
	\node[draw, circle,inner sep=-0.0pt, right=\arrowLen of sld] (noise) {$\mathbf{+}$};

    \node[block, rounded corners, right=\arrowLen of noise, minimum width=\minWid] (rxfilter) {RRC};
    \node[block, rounded corners, minimum width=\minWid, minimum height=\minHei, right=\arrowLen+0.2cm of rxfilter] (equalizer) {SNN};

	\draw[-{Latex[length=2mm]}, thick] (transmitter) -- node[midway, above, xshift=-0.1cm] {$\bm{x}$} (txfilter);
	\draw[-{Latex[length=2mm]}, thick] (txfilter) -- (channel);
	\draw[-{Latex[length=2mm]}, thick] (channel) -- node[midway, above] {$\bm{\tilde{x}}$} (sld);
	\draw[-{Latex[length=2mm]}, thick] (sld) -- (noise);
	\draw[{Latex[length=2mm]}-, thick] (noise) -- node[midway, right, yshift=+0.2cm] {$\bm{n}$} +(0,1.5*\arrowLen);
    \draw[-{Latex[length=2mm]}, thick] (noise) -- (rxfilter);
	\draw[-{Latex[length=2mm]}, thick] (rxfilter.east) -- node[midway, above] {$\bm{y}$}  (equalizer.west);
    \draw[-{Latex[length=2mm]}, thick] (equalizer.east) -- node[midway, above, xshift=+0.2cm] {$\bm{\hat{x}}$} +(\arrowLen,0);
    \end{pgfonlayer}

    \node[draw, RPTU_BlueGray, thick, dotted, rounded corners, inner xsep=0.1cm, inner ysep=0.15cm, fit=(transmitter)(txfilter)] (transmitter_highlight) {};
    \node[fill=white, yshift=-0.5mm] at (transmitter_highlight.south) {\textcolor{RPTU_BlueGray}{\footnotesize Transmitter}};

    \node[draw, RPTU_Violett, thick, dotted, rounded corners, inner xsep=0.1cm, inner ysep=0.15cm, fit=(channel) (sld) (noise)] (total_channel) {};
    \node[fill=white, yshift=-0.5mm] at (total_channel.south) {\textcolor{RPTU_Violett}{\footnotesize Optical Fiber Channel}};

    \node[draw, RPTU_GreenGray, thick, dotted, rounded corners, inner xsep=0.1cm, inner ysep=0.15cm, fit=(equalizer)(rxfilter)] (receiver) {};
    \node[fill=white, yshift=-0.5mm] at (receiver.south) {\textcolor{RPTU_GreenGray}{\footnotesize Receiver}};

\end{tikzpicture}
 \vspace{-0.5cm}
	\caption{Model of the optical communication chain. Data is sent from a transmitter over an optical fiber channel with \ac{cd} and square-law detection to a receiver including the \ac{snn}-based equalizer.}
	\label{fig:channel_model}
  \end{center}
\end{figure}
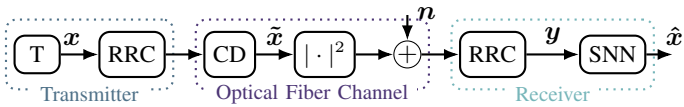

As transmission link we select the optical fiber channel with \ac{imdd} and \ac{pam} proposed by \cite{arnold2022} which is also used in \cite{vonbank2023spiking}. As shown in Fig. \ref{fig:channel_model} the transmitted symbols $\mathbf{x}$ are upsampled by a factor of two and convolved with a \ac{rrc} pulse shaping filter with a roll-off factor of \num{0.2} at the transmitter. The simulated optical fiber channel applies \ac{cd} with a fiber dispersion coefficient of \SI{17}{\CD} followed by square-law detection which distorts the signal nonlinearly. Afterwards, the signal is superimposed by \ac{awgn} with zero mean and unit variance. At receiver side, the signal is \ac{rrc}-filtered and downsampled. Finally, the \ac{snn}-based equalizer is applied to compensate for the distortions of the channel by producing an estimate of the transmitted symbol, given as $\bm{\tilde{x}}$. 
In particular, the channel has the following parameters: a fiber length of \SI{5}{\kilo \meter}, a data rate of \SI{50}{\giga Bd}, and a wavelength of \SI{1550}{\nano \meter}. Further, \ac{pam}-4 is applied where the symbols $\mathcal{C} = \{ 0, 1, \sqrt{2}, \sqrt{3}\}$ are transmitted with Gray mapping. This configuration corresponds to the channel B introduced in \cite{vonbank2023spiking}. 

Due to \ac{cd} in combination with square-law detection, the transmitted signal is distorted nonlinearly. Those distortions can't be compensated by conventional linear equalizers. This motivates for the use of either traditional \acp{dfe} or modern \ac{nn}-based equalization approach due to their ability to compensate for nonlinear impairments~\cite{ney2023}.

\section{Spiking Neural Networks}
\Acp{snn} are the next generation of \acp{nn}\cite{9585245}. The main difference between \acp{snn} and \acp{ann} is that their functionality is more similar to that of the human brain, in which neurons communicate by action potentials. On digital devices, action potentials can be modelled as binary signals. This simplifies the hardware implementation and increases the energy efficiency. Furthermore, biological neurons work in an event-based way; as a result, they perform calculations only if they receive an input spike. Consequently \acp{snn} are expected to be more energy efficient than similarly sized \acp{ann}. 

\acp{snn} are time-dependent, in other words, for each input sample the network runs for a specific number of iterations or time steps. The various neurons charge or discharge at each time step according to their internal parameters and the state of their input. 

Multiple neuron models were developed to mimic the biological neurons. They range from accurate or biologically plausible such as the Hodgkin-Huxley\cite{Hodgkin} model to simple ones; that do not support most of the biological features but have lower computational complexity such as the \ac{lif} model.

\subsection{\Acl{lif} model}
The \ac{lif} model is a commonly used neuron model. A recurrent version of the model is described by (\ref{eq:lif_v_equation}) and (\ref{eq:lif_i_equation}) \cite{neftci2019surrogate}:
\begin{equation}
\label{eq:lif_v_equation}
\frac{dv(t)}{dt} = -\frac{(v(t)-v_r)+ri(t)}{\tau_m}+\Theta(v(t)-v{_\mathrm{th}})(v_r-v{_\mathrm{th}})
\end{equation}

\begin{equation}
\label{eq:lif_i_equation}
\frac{di(t)}{dt} = -\frac{i(t)}{\tau_s}+\sum_jw_js_j(t)+\sum_jv_js_rj(t)
\end{equation}

\begin{figure}
	\centering
	\tikzsetnextfilename{tikz_lif_graph}
\begin{tikzpicture}

  \begin{axis}[
            legend style={font=\footnotesize},
            legend pos=north east,
            xlabel={Time (ms)},
            ylabel=,
            xmin=0,
            xmax=1600,
            width=\columnwidth,
            height=0.8*\columnwidth,
            ]

    \addplot [RPTU_BlueGray, solid,line width=1pt] table [x=time,y=voltage]{figures/data/LIF_graph.txt}; \addlegendentry{Voltage};
    \addplot [RPTU_Orange, solid,line width=1pt] table [x=time,y=current]{figures/data/LIF_graph.txt}; \addlegendentry{Current};
    \addplot+ [Navy_Blue, only marks, mark=10-pointed star, mark size=3pt] coordinates{(20,0) (200,0) (250,0) (300,0) (800,0) (850,0) (900,0)}; \addlegendentry{Input Spikes};
    \addplot+ [RPTU_Red, only marks, mark size=6pt] coordinates{(314,1) (936,1)}; \addlegendentry{Output Spikes};
    
    \draw [solid, RPTU_Pink] (0,1) -- (1600,1);
    \draw [black,-, thick] (axis cs:100,1) node[anchor=south] {$v_\mathrm{th}$};

        \draw [solid, black] (0,0) -- (1600,0);
    \draw [black,-, thick] (axis cs:100,0) node[anchor=south] {$v_\mathrm{r}$};
    \end{axis}

\end{tikzpicture}
        \caption{Simulation of the \ac{lif} model. Voltage and current increase with the arrival of new spikes and an output spike is generated if the voltage reaches $v_\mathrm{th}$, after which it's set to $v_\mathrm{r}$}
        \label{fig:lif-behaviour}
\end{figure}
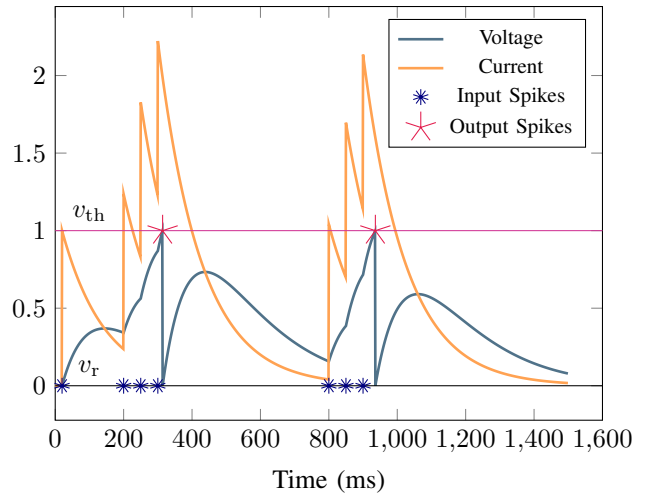

Where $v(t)$ is the voltage or membrane potential, $v_r$ is the reset voltage, $r$ is the input resistance, $i(t)$ is the input current, $\Theta(\cdot)$ denotes the Heaviside function, $v{_\mathrm{th}}$ is the threshold voltage, $\tau_m$ is the voltage-time constant, $\tau_s$ is the input time constant, $w_j$ is the input weight, $s_j(t)$ is the input from the previous layer, $v_j$ is the weight of the recurrent connection and $s_rj(t)$ is the input from the recurrent connection. Fig. \ref{fig:lif-behaviour} illustrates the working principle of the \ac{lif} model. Once an input spike is received the voltage increases and then starts to decay; $\tau_m$ controls the decaying speed. The input current also decays with a factor of $\tau_s$. If multiple consecutive spikes are received by the neuron the voltage increases till it exceeds the threshold; then an output spike is generated and the voltage is set to the reset voltage.

\subsection{SNN structure}
\begin{figure}[t]
\centering
\includegraphics[width=0.95\columnwidth]{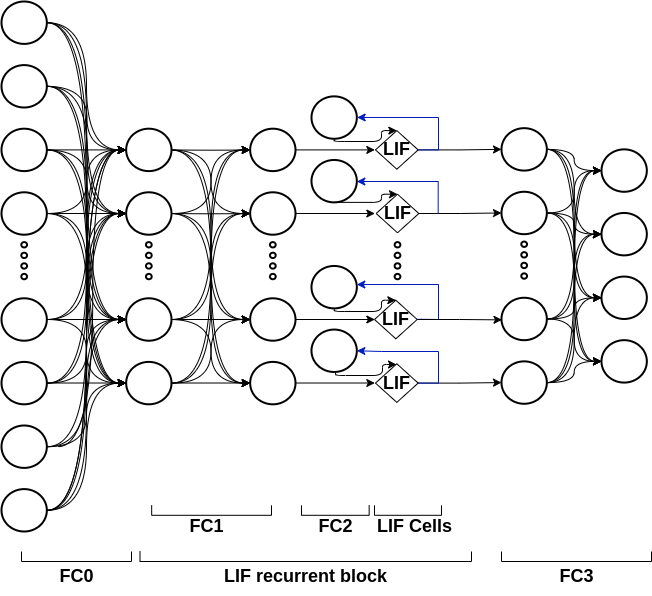}
\caption{The architecture of the reference topology, where FC0 is the linear input layer, FC1 is the \ac{lif} linear input layer, FC2 is the \ac{lif} recurrent layer, \ac{lif} Cells are \ac{lif} activation function and FC3 is the linear output layer. Blue arrows represent the recurrent connections.}
\label{fig:snn-dfe-arch}
\end{figure}

This work is based on the SNN-DFE network presented in \cite{vonbank2023spiking}, referred to later as the reference topology. The structure of the reference topology is illustrated in Fig. \ref{fig:snn-dfe-arch}, it consists of three components: the linear input layer, the \ac{lif} recurrent block, and the linear output layer. The \ac{lif} recurrent block consists of three layers: a linear input layer, a recurrent linear layer, and an activation layer consisting of the \ac{lif} neurons. The topology processes the input in the following way using ten time steps. For the first time step the input is encoded using the encoding mechanism introduced in \cite{bansbach2023spiking} and passed to the network. For the remaining nine steps zeros are passed as input. On the output side, the output of each neuron is accumulated over the time steps and the output with the highest value is considered as the final decision.

\section{Design Space Exploration}
\label{sec:design_space_exploration}

Due to the huge number of different hyperparameters of \acp{nn} like the depth of the network, the type of layers, the activation function, the number of neurons, and many more, there is a huge design space of various \ac{nn} topologies. This is the case for \acp{snn} as well as for \acp{ann}. When considering only the algorithmic side, a detailed design space exploration is not necessarily required since a huge \ac{nn} topology often provides sufficient computational complexity to find an accurate mapping between input and output. However, for efficient hardware implementation, it is crucial to consider the complexity of the \ac{nn} resulting in a trade-off between communication performance and computational complexity. This is especially critical for resource-constraint platforms like \acp{fpga}.

\subsection{Design Parameters and Evaluation Metrics}
For the application of \ac{nn}-based equalization, the algorithmic performance is expressed as \ac{ber} while the computational complexity is given in terms of \ac{mac} operations. For our \ac{snn} topology template, the \ac{mac} operations per output symbol can be calculated as: 
\begin{equation}
\label{eq:mac_ops}
    \mathrm{MAC}_\mathrm{SNN} = N_\mathrm{H} \cdot (N_\mathrm{I} + 2 \cdot N_\mathrm{H} + 4) \cdot T \; ,
\end{equation}

where $N_\mathrm{H}$ corresponds to the number of neurons in the hidden layer, $N_\mathrm{I}$ corresponds to the number of inputs, and $T$ gives the number of \ac{snn} time steps.  
For the reference topology, this results in \num{329600} \ac{mac} operations per output symbol. When comparing it to the \ac{ann} proposed in \cite{vonbank2023spiking} with \num{3600} \ac{mac} operations, which is two orders of magnitude lower, it becomes clear that the \ac{snn} is by far too complex to achieve competitive hardware performance. Thus a detailed design space exploration of the \ac{snn} topology is required to provide a fair comparison between the \ac{ann} and the \ac{snn} regarding implementation efficiency.

Since the general structure of our \ac{snn} topology is already given by the reference topology, we can constrain our design space of the \ac{snn} topology to the following hyperparameters: the input size, the number of hidden neurons and the time steps. 

Examining the reference topology one can notice that the input layer has the most \ac{mac} operations. The number of \ac{mac} operations of the input layer is determined by the number of inputs and hidden neurons. Thus, reducing those is crucial for designing a low-complex \ac{snn} topology. The number of inputs depends on the tap count which is divided into 3 sets of values. The first set is the last $\lfloor \frac{n_\mathrm{tap}}{2} \rfloor$ received symbols, the second set is the corresponding $\lfloor \frac{n_\mathrm{tap}}{2} \rfloor$ estimated symbols, and the final set is the currently received symbol. Received symbols are encoded with 8 bits and estimations are encoded with $2^m$ bits. Equation (\ref{eq:in_features_tap_count}) describes the relationship between tap count and the number of input neurons.

\begin{equation}
\label{eq:in_features_tap_count}
N_\mathrm{I} = 8 \cdot (\lfloor \frac{n_\mathrm{tap}}{2} \rfloor+1) + 2^m \cdot (\lfloor \frac{n_\mathrm{tap}}{2} \rfloor)
\end{equation}
where $n_\mathrm{tap}$ is the number of equalizer taps and $m$ is the modulation order. 

The number of hidden neurons has an immense impact on the topology's computational complexity as it controls the size of all the layers, from (\ref{eq:mac_ops}) it can be seen that the complexity grows quadratically with the number of hidden neurons. 
Finally, the number of time steps directly dictates how many times the topology has to run to generate one output. Therefore, $\mathrm{MAC}_\mathrm{SNN}$ depends linearly on the number of time steps.

\subsection{Training and Simulation Setup}
\label{sec:training_simulation_setup}
The different topologies are trained using PyTorch\cite{paszke2019pytorch} and Norse\cite{norse2021} on an NVIDIA A30 graphics card, with the configuration defined in \cite{vonbank2023spiking}. The networks are trained for $5$ epochs using $10000$ batches with a batch size of $200000$ and a learning rate of $0.001$. Furthermore, the cross entropy loss function is used with the Adam optimizer. All configurations are trained for a \ac{snr} of $17$ and tested for \acp{snr} of 12 to 21. To guarantee a fair comparison, deterministic PyTorch settings are used during training and testing.

\subsection{Design Space Exploration Decisions}
\label{dse_res}
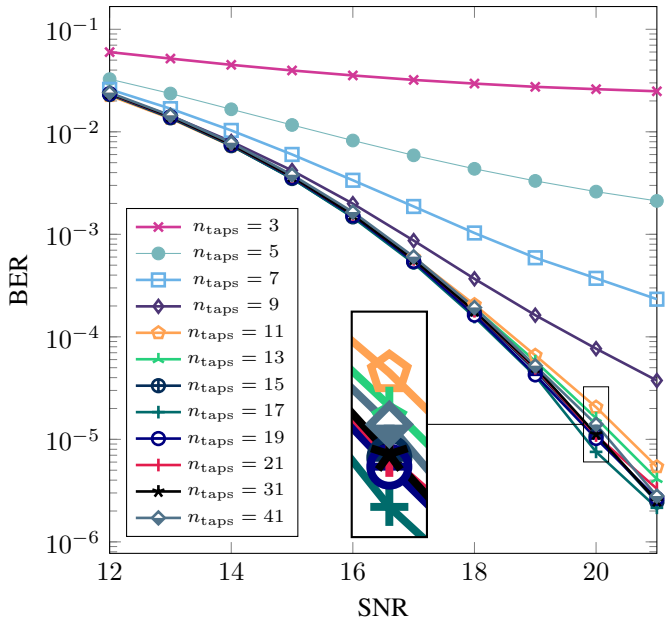
\begin{figure}
	\centering
	\tikzsetnextfilename{tikz_ber_vs_snr_tap_count}
\begin{tikzpicture}[spy using outlines={rectangle, magnification=3, connect spies}]
	\begin{axis}[
            ymode=log, 
            xmin=12,
            xmax=21,
            legend style={font=\scriptsize},
            legend pos=south west,
            xlabel={SNR},
            ylabel={BER},
            width=\columnwidth,
            height=\columnwidth,
            ]



    \addplot [RPTU_Pink, solid,line width=1pt,  mark=x, mark options={scale=1.2,solid}] table [x=EbN0,y=3]{figures/data/ber_per_snr_diff_tap_count.txt}; \addlegendentry{$n_\mathrm{taps}=3$};
    
    \addplot [RPTU_GreenGray, solid, width=1pt,mark=*, mark options={scale=1.2,solid}] table [x=EbN0,y=5]{figures/data/ber_per_snr_diff_tap_count.txt}; \addlegendentry{$n_\mathrm{taps}=5$};

    \addplot [RPTU_LightBlue, solid, line width=1pt,mark=square, mark options={scale=1.2,solid}] table [x=EbN0,y=7]{figures/data/ber_per_snr_diff_tap_count.txt}; \addlegendentry{$n_\mathrm{taps}=7$};

    \addplot [RPTU_Violett , solid, line width=1pt,mark=diamond, mark options={scale=1.2,solid}] table [x=EbN0,y=9]{figures/data/ber_per_snr_diff_tap_count.txt}; \addlegendentry{$n_\mathrm{taps}=9$};

    \addplot [RPTU_Orange, solid,line width=1pt,mark=pentagon, mark options={scale=1.2,solid}] table [x=EbN0,y=11]{figures/data/ber_per_snr_diff_tap_count.txt}; \addlegendentry{$n_\mathrm{taps}=11$};

    \addplot [RPTU_LightGreen, solid,line width=1pt,mark=Mercedes star, mark options={scale=1.2,solid}] table [x=EbN0,y=13]{figures/data/ber_per_snr_diff_tap_count.txt}; \addlegendentry{$n_\mathrm{taps}=13$};

    \addplot [RPTU_DarkBlue, solid,line width=1pt,mark=oplus, mark options={scale=1.2,solid}] table [x=EbN0,y=15]{figures/data/ber_per_snr_diff_tap_count.txt}; \addlegendentry{$n_\mathrm{taps}=15$};

    \addplot [RPTU_DarkGreen, solid,line width=1pt,mark=+, mark options={scale=1.2,solid}] table [x=EbN0,y=17]{figures/data/ber_per_snr_diff_tap_count.txt}; \addlegendentry{$n_\mathrm{taps}=17$};

    \addplot [Navy_Blue, solid,line width=1pt,mark=halfcircle, mark options={scale=1.2,solid}] table [x=EbN0,y=19]{figures/data/ber_per_snr_diff_tap_count.txt}; \addlegendentry{$n_\mathrm{taps}=19$};

    \addplot [RPTU_Red, solid,line width=1pt,mark=|, mark options={scale=1.2,solid}] table [x=EbN0,y=21]{figures/data/ber_per_snr_diff_tap_count.txt}; \addlegendentry{$n_\mathrm{taps}=21$};

    \addplot [RPTU_Black, solid,line width=1pt,mark=star, mark options={scale=1.2,solid}] table [x=EbN0,y=31]{figures/data/ber_per_snr_diff_tap_count.txt}; \addlegendentry{$n_\mathrm{taps}=31$};

    \addplot [RPTU_BlueGray, solid,line width=1pt,mark=halfsquare*, mark options={scale=1.2,solid}] table [x=EbN0,y=41]{figures/data/ber_per_snr_diff_tap_count.txt}; \addlegendentry{$n_\mathrm{taps}=41$};

    \coordinate (spypoint) at (axis cs:20,1.40E-05);
    \coordinate (magnifyglass) at (axis cs:16.6,1.40E-05);
  
    \end{axis}
    \spy [black, height=3cm, width=1cm] on (spypoint) in node[fill=white] at (magnifyglass);
\end{tikzpicture}
	\caption{Communication performance of different $n_\mathrm{tap}$}
	\label{fig:ber_vs_snr_tap_count}
\end{figure}

In the design space exploration, first, the number of input neurons is optimized by determining an appropriate value for $n_\mathrm{tap}$ by training and testing multiple variants of the reference topology. Fig.  \ref{fig:ber_vs_snr_tap_count} presents a selected set of the results from which it's evident that $n_\mathrm{tap} = 17$ provides the best performance for most SNRs compared to $n_\mathrm{tap} = 41$ which is used in the reference topology. According to (\ref{eq:in_features_tap_count}),  $n_\mathrm{tap} = 17$ reduces the number of input neurons from $248$ to $104$, consequently this leads to a \SI{34.95}{\percent} reduction in the total number of \ac{mac} operations compared to the reference topology.

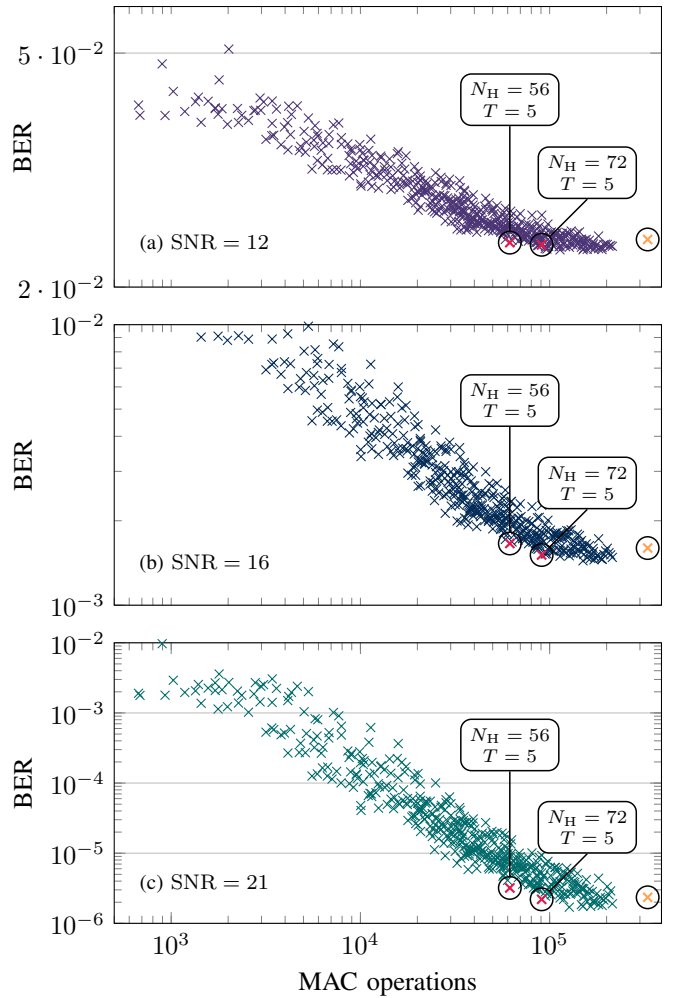
\begin{figure}[htbp]
	\centering
	\tikzsetnextfilename{optuna_results}

\begin{tikzpicture}
\pgfplotssetlayers{background,background_1,background_2,background_3,main,foreground_6,foreground_5,foreground_4,foreground_3,foreground_2,foreground_1,foreground}

    \begin{groupplot}[
        group style={
            group name=optuna results group plot,
            group size=1 by 3,
            vertical sep=5mm
        },
        xmode=log,
        ymode=log, 
        ymajorgrids,
        xmin=500,
        xmax=390000,
        legend pos=north east
    ]

    \nextgroupplot[
        width=\columnwidth,
        height=0.6*\columnwidth,
        xmajorticks=false,
        ymin=2e-2,
        ymax=6e-2,
        ytick={0.02, 0.05},
        yticklabels={$2\cdot10^{-2}$, $5\cdot10^{-2}$},
        ylabel={BER},
        ylabel shift = -12pt
    ]

    \addplot [only marks, mark=x, color=RPTU_Violett, mark options={scale=1.2}] 
        table [x=MAC, y=BER] {figures/data/optuna_results_snr12.txt};

    \draw [black,-, thick] (axis cs:1500,0.022) node[anchor=south] {\footnotesize (a) $\mathrm{SNR}=12$};

    \coordinate (H72_S5_SNR12) at (axis cs:90720, 0.0236);
    \coordinate (H56_S5_SNR12) at (axis cs:61600, 0.0238);
    \coordinate (FSNN_SNR12) at (axis cs:329600, 0.0240970999);


    \begin{pgfonlayer}{foreground}

    \draw[line width=0.2mm, black] (H72_S5_SNR12) circle [x radius=0.15cm, y radius=0.15cm];
    \addplot [only marks, mark=x, mark options={RPTU_Red, scale=1.2, thick}] coordinates {(90720, 0.0236)};
    \draw [black, line width=0.2mm] ($(H72_S5_SNR12) + (0.107cm, 0.107cm)$) -- ++(0.5cm, 0.5cm);
    \node[block, draw=black, rounded corners, line width=0.2mm, anchor=south, font=\scriptsize, fill=white] at ($(H72_S5_SNR12) + (0.607cm, 0.607cm)$) {$N_\mathrm{H}=72$\\ $T=5$};    

    \draw[line width=0.2mm, black, on layer=foreground_1] (H56_S5_SNR12) circle [x radius=0.15cm, y radius=0.15cm];
    \addplot [only marks, mark=x, mark options={RPTU_Red, scale=1.2, thick}]  coordinates {(61600, 0.0238)};
    \draw [black, line width=0.2mm] ($(H56_S5_SNR12) + (0cm, 0.15cm)$) -- ++(0cm, 1.4cm);
    \node[block, draw=black, rounded corners, line width=0.2mm, anchor=south, font=\scriptsize, fill=white] at ($(H56_S5_SNR12) + (0cm, 1.55cm)$) {$N_\mathrm{H}=56$\\ $T=5$};

    \draw[line width=0.2mm, black, on layer=foreground_1] (FSNN_SNR12) circle [x radius=0.15cm, y radius=0.15cm];
    \addplot [only marks, mark=x, mark options={RPTU_Orange, scale=1.2, thick}]  coordinates {(329600, 0.0240970999)};

    \end{pgfonlayer}

    \nextgroupplot[
        width=\columnwidth,
        height=0.6*\columnwidth,
        xmajorticks=false,
        ymin=1e-3,
        ymax=1e-2,
        ylabel={BER},
    ]
    \addplot [only marks, mark=x, color=RPTU_DarkBlue, mark options={scale=1.2}] 
        table [x=MAC, y=BER] {figures/data/optuna_results_snr16.txt};

    \draw [black,-, thick] (axis cs:1500,0.0012) node[anchor=south] {\footnotesize (b) $\mathrm{SNR}=16$};

    \coordinate (H72_S5_SNR16) at (axis cs:90720, 0.00151);
    \coordinate (H56_S5_SNR16) at (axis cs:61600, 0.00166);
    \coordinate (FSNN_SNR16) at (axis cs:329600, 0.00159879995);

    \begin{pgfonlayer}{foreground}

    \addplot [only marks, mark=x, mark options={RPTU_Red, scale=1.2, thick}] coordinates {(90720, 0.00151)};
    \draw[line width=0.2mm, black] (H72_S5_SNR16) circle [x radius=0.15cm, y radius=0.15cm];
    \draw [black, line width=0.2mm] ($(H72_S5_SNR16) + (0.107cm, 0.107cm)$) -- ++(0.5cm, 0.5cm);
    \node[block, draw=black, rounded corners, line width=0.2mm, anchor=south, font=\scriptsize, fill=white] at ($(H72_S5_SNR16) + (0.607cm, 0.607cm)$) {$N_\mathrm{H}=72$\\ $T=5$};    

    \addplot [only marks, mark=x, mark options={RPTU_Red, scale=1.2, thick}] coordinates {(61600, 0.00166)};
    \draw[line width=0.2mm, black] (H56_S5_SNR16) circle [x radius=0.15cm, y radius=0.15cm];
    \draw [black, line width=0.2mm] ($(H56_S5_SNR16) + (0cm, 0.15cm)$) -- ++(0cm, 1.4cm);
    \node[block, draw=black, rounded corners, line width=0.2mm, anchor=south, font=\scriptsize, fill=white] at ($(H56_S5_SNR16) + (0cm, 1.55cm)$) {$N_\mathrm{H}=56$\\ $T=5$};

    \draw[line width=0.2mm, black, on layer=foreground_1] (FSNN_SNR16) circle [x radius=0.15cm, y radius=0.15cm];
    \addplot [only marks, mark=x, mark options={RPTU_Orange, scale=1.2, thick}]  coordinates {(329600, 0.00159879995)};



    \end{pgfonlayer}

    \begin{pgfonlayer}{foreground_1}

    \end{pgfonlayer}

    \nextgroupplot[
        width=\columnwidth,
        height=0.6*\columnwidth,
        ymin=1e-6,
        ymax=1e-2,
        ylabel={BER},
        xlabel={MAC operations}
    ]
    \addplot [only marks, mark=x, color=RPTU_DarkGreen, mark options={scale=1.2}] 
        table [x=MAC, y=BER] {figures/data/optuna_results_snr21.txt};

    \draw [black,-, thick] (axis cs:1500,2e-6) node[anchor=south] {\footnotesize (c) $\mathrm{SNR}=21$};

    \coordinate (H72_S5_SNR21) at (axis cs:90720, 2.2e-6);
    \coordinate (H56_S5_SNR21) at (axis cs:61600, 3.2e-6);
    \coordinate (FSNN_SNR21)   at (axis cs:329600, 0.00000235);

    \begin{pgfonlayer}{foreground}

    \addplot [only marks, mark=x, mark options={RPTU_Red, scale=1.2, thick}] coordinates {(90720, 2.2e-6)};
    \draw[line width=0.2mm, black] (H72_S5_SNR21) circle [x radius=0.15cm, y radius=0.15cm];
    \draw [black, line width=0.2mm] ($(H72_S5_SNR21) + (0.107cm, 0.107cm)$) -- ++(0.5cm, 0.5cm);
    \node[block, draw=black, rounded corners, line width=0.2mm, anchor=south, font=\scriptsize, fill=white] at ($(H72_S5_SNR21) + (0.607cm, 0.607cm)$) {$N_\mathrm{H}=72$\\ $T=5$};    

    \addplot [only marks, mark=x, mark options={RPTU_Red, scale=1.2, thick}] coordinates {(61600, 3.2e-6)};
    \draw[line width=0.2mm, black] (H56_S5_SNR21) circle [x radius=0.15cm, y radius=0.15cm];
    \draw [black, line width=0.2mm] ($(H56_S5_SNR21) + (0cm, 0.15cm)$) -- ++(0cm, 1.4cm);
    \node[block, draw=black, rounded corners, line width=0.2mm, anchor=south, font=\scriptsize, fill=white] at ($(H56_S5_SNR21) + (0cm, 1.55cm)$) {$N_\mathrm{H}=56$\\ $T=5$};    

    \draw[line width=0.2mm, black, on layer=foreground_1] (FSNN_SNR21) circle [x radius=0.15cm, y radius=0.15cm];
    \addplot [only marks, mark=x, mark options={RPTU_Orange, scale=1.2, thick}]  coordinates {(329600, 0.00000235)};



    \end{pgfonlayer}

    \end{groupplot}

\end{tikzpicture}
	\caption{Design space exploration results for number of hidden neurons and time steps for $n_\mathrm{tap}=17$, the orange mark represents the reference topology.}
	\label{fig:optuna_results}
\end{figure}

The remaining design parameters are the number of hidden neurons and the time steps. The reference topology has $80$ hidden neurons and $10$ time steps. We utilised the hyper-parameters optimization framework Optuna \cite{optuna_2019} to search the ranges of $1-80$ hidden neurons and $1-10$ time steps for promising configurations. After training and testing each configuration, the results are grouped by \ac{snr} and plotted to illustrate the communication performance versus the total number of \ac{mac} operations. Fig. \ref{fig:optuna_results} presents the results of 3 \acp{snr}, namely 12, 16 and 21. The orange marks represent the reference topology with $n_\mathrm{tap} = 41$, 80 hidden neurons and 10 time steps. After analyzing the resulting plots, two configurations stand out as they outperform the reference topology at a lower cost, namely {$N_H=56, T=5$} and {$N_H=72, T=5$}.
In the following, we will refer to those topologies as $\mathrm{SNN}_{56}$ and $\mathrm{SNN}_{72}$ and to the reference topology as $\mathrm{SNN}_\mathrm{ref}$.

\section{Hardware Architecture}
The topology is implemented using \ac{hls} under the Vitis-HLS 2023.2 environment. First, we extend Xilinx's Finn-HLSlib to support a recurrent \ac{lif} layer. Then the topology is implemented in a templated way to ease the exploration of the design space. Layers are defined using templated arbitrary precision data types. The topology is implemented in a streaming way in which each layer is implemented as a stage in a pipeline, layers are connected via streams. Each layer allows for a variable degree of input and output parallelism for exploration of the trade-off between power and throughput. Our target platform is the Xilinx ZCU104 evaluation board. It features the Xilinx XCZU7EV including a quad-core Arm Cortex A53 as a \ac{cpu} and a Xilinx \ac{fpga}. The processing starts on the \ac{cpu} by encoding the input symbols and transferring them to the shared memory. The \ac{pl} reads the encoded symbols from the shared memory through AXI-Streams, processes it and writes the results back.

The original Norse\cite{norse2021} implementation has $\tau_m$ and $\tau_s$ set to 100 and 200 respectively. For an efficient \ac{fpga} implementation we replace those values with $125$ and $250$ respectively. This way we could replace the complex multiply operation with a low complexity shift operation.

We convert the 32-bit floating-point parameters used in the Python environment to arbitrary-width fixed-point format. The process of converting the values to fixed-point is referred to as quantization. There are two types of quantization, namely \ac{qat} and \ac{ptq}. In this work \ac{qat} is utilized as it results in higher accuracy compared to \ac{ptq}. Quantized linear layers from Brevitas \cite{brevitas} are used. A quantized recurrent \ac{lif} layer is developed based on the LIF implementation of Norse. The input symbols are limited to {-1, 0, 1} and can therefore be represented with 4 bits. In contrast, for weights, bias, voltage and current, the bit width is chosen based on a trade-off analysis of complexity and communication performance.

\section{Results}
\label{sec:results}
\subsection{Communication Performance}
Following the design space exploration flow described in section \ref{sec:design_space_exploration} we get an optimized version of the reference topology that uses $n_\mathrm{tap} = 17$, $T=5$, $\tau_m = 125$, $\tau_s = 250$  and $N_H=56$ or $72$ hidden neurons. 

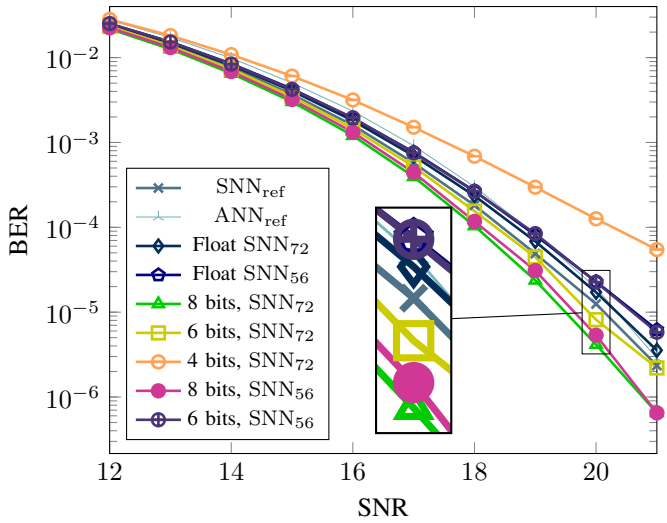
\begin{figure}
	\centering
	\tikzsetnextfilename{quant_snn_performance}
\usetikzlibrary{spy}
\begin{tikzpicture}[spy using outlines={rectangle, magnification=2.7, connect spies}]
	\begin{axis}[
            ymode=log, 
            xmin=12,
            xmax=21,
            ymax=0.04,
            legend style={font=\footnotesize},
            legend pos=south west,
            xlabel={SNR},
            ylabel={BER},
            width=\columnwidth,
            height=\columnwidth*0.85,
            ]

    \addplot [RPTU_BlueGray, solid,line width=1pt,  mark=x, mark options={scale=1.2,solid}] table [x=EbN0,y=Float_SNN]{figures/data/snn_vs_ann_performance.txt}; \addlegendentry{$\mathrm{SNN}_\mathrm{ref}$};

    \addplot [RPTU_GreenGray, solid, width=1pt,mark=Mercedes star, mark options={scale=1.2,solid}] table [x=EbN0,y=Float_ANN]{figures/data/snn_vs_ann_performance.txt}; \addlegendentry{$\mathrm{ANN}_\mathrm{ref}$};

    \addplot [RPTU_DarkBlue , solid, line width=1pt,mark=diamond, mark options={scale=1.2,solid}] table [x=EbN0,y=SNN_H72_S5_SH]{figures/data/snn_vs_ann_performance.txt}; \addlegendentry{Float $\mathrm{SNN}_{72}$};

    \addplot [Navy_Blue, solid,line width=1pt,mark=pentagon, mark options={scale=1.2,solid}] table [x=EbN0,y=SNN_H56_S5_SH]{figures/data/snn_vs_ann_performance.txt}; 	\addlegendentry{Float $\mathrm{SNN}_{56}$};

    \addplot [green, solid, line width=1pt,mark=triangle, mark options={scale=1.2,solid}] table [x=EbN0,y=8bits]{figures/data/snn_quant_performance.txt}; \addlegendentry{8 bits, $\mathrm{SNN}_{72}$};

    \addplot [yellow , solid, line width=1pt,mark=square, mark options={scale=1.2,solid}] table [x=EbN0,y=6bits]{figures/data/snn_quant_performance.txt}; \addlegendentry{6 bits, $\mathrm{SNN}_{72}$};

    \addplot [RPTU_Orange, solid,line width=1pt,mark=halfcircle, mark options={scale=1.2,solid}] table [x=EbN0,y=4bits]{figures/data/snn_quant_performance.txt}; \addlegendentry{4 bits, $\mathrm{SNN}_{72}$};

    \addplot [RPTU_Pink, solid,line width=1pt,mark=*, mark options={scale=1.2,solid}] table [x=EbN0,y=QH568bits]{figures/data/snn_quant_performance.txt}; \addlegendentry{8 bits, $\mathrm{SNN}_{56}$};

    \addplot [RPTU_Violett, solid,line width=1pt,mark=oplus, mark options={scale=1.2,solid}] table [x=EbN0,y=QH566bits]{figures/data/snn_quant_performance.txt}; \addlegendentry{6 bits, $\mathrm{SNN}_{56}$};
    
    \coordinate (spypoint) at (axis cs:20,0.00001);
    \coordinate (magnifyglass) at (axis cs:17,0.000008);
    \end{axis}
    \spy [black, height=3cm, width=1cm] on (spypoint) in node[fill=white] at (magnifyglass);
\end{tikzpicture}
	\caption{The communication performance of the reference floating point topology compared to the communication performance of $N_\mathrm{H}=72$ and $N_\mathrm{H}=56$ and their quantized variants.}
	\label{fig:ann_snn_performance}
\end{figure}

The \ac{ber} of $\mathrm{SNN}_{72}$ is higher than that of $\mathrm{SNN}_\mathrm{ref}$ by an average of \SI{21.23}{\percent}. However, after applying 8 bits \ac{qat}, a considerable improvement in the communication performance by \SI{34.45}{\percent} compared to $\mathrm{SNN}_\mathrm{ref}$ and by \SI{52.03}{\percent} compared to $\mathrm{ANN}_\mathrm{ref}$ is achieved; which outperformed all other experiments we investigated as illustrated in Fig. \ref{fig:ann_snn_performance}.
Reducing the bit width to 6 bits gives a modest improvement by \SI{10.06}{\percent} on average compared to $\mathrm{SNN}_\mathrm{ref}$. Additionally, 4 bits were implemented but as illustrated in Fig. \ref{fig:ann_snn_performance}, it results in the lowest performance. From a computational complexity perspective, $\mathrm{SNN}_{72}$ has $3.63\times$ less \ac{mac} operations than $\mathrm{SNN}_\mathrm{ref}$.

The communication performance of $\mathrm{SNN}_{56}$ is also presented in Fig. \ref{fig:ann_snn_performance}. The \ac{ber} of the float variant increases by \SI{45.03}{\percent} on average, however, its 8 bits variant achieves a \ac{ber} reduction of \SI{25.11}{\percent} as compared to the $\mathrm{SNN}_\mathrm{ref}$. Unlike the $N_\mathrm{H}=72$ version, the communication performance of the 6 bits variant of this version is lower by \SI{43.3}{\percent} on average in comparison to the reference topology.
$\mathrm{SNN}_{56}$ has $5.35\times$ less \ac{mac} operations than $\mathrm{SNN}_\mathrm{ref}$ and $1.47\times$ less than $\mathrm{SNN}_{72}$. 

\subsection{FPGA Implementation}
After analyzing the communication performance results we implemented the two topologies namely 8 bits, $\mathrm{SNN}_{72}$ and 8 bits, $\mathrm{SNN}_{56}$ in addition to the reference \ac{ann} and \ac{snn} from \cite{vonbank2023spiking}. 
To make an objective and fair comparison we adjust the input and output parallelism of the topologies such that they have roughly the same throughput. Table \ref{tab:implementation_results} presents the results. We run each topology for $100K$ cycles, in each cycle $1000$ samples are processed in burst mode. We measured the latency and power of the \ac{pl} and averaged them over the total number of samples. The power of the \ac{pl} is composed of two components, a static component and a dynamic component. Static power is the power consumed by the \ac{pl} without stimulating any input, while dynamic power is consumed while stimulating input. Comparing $\mathrm{ANN}_\mathrm{ref}$ with $\mathrm{SNN}_\mathrm{ref}$ we see that $\mathrm{SNN}_\mathrm{ref}$ requires $2.86\times$ more power while its dynamic energy is higher by a factor of $9.52$. 
On the other side, $\mathrm{SNN}_{72}$ requires a total power that is higher than $\mathrm{ANN}_\mathrm{ref}$ by $1.16\times$ while having almost the same dynamic energy. With respect to the energy efficiency we see that the unoptimized $\mathrm{SNN}_\mathrm{ref}$ already has $9.6\times$ better energy efficiency compared to $\mathrm{ANN}_\mathrm{ref}$; while $\mathrm{SNN}_{72}$ pushes this number even further to $25\times$.

In summary, our proposed versions have considerable gains in communication performance at an order of magnitude higher energy efficiency than $\mathrm{ANN}_\mathrm{ref}$.

\begin{table}
\centering

\caption{Hardware implementation results}
\label{tab:implementation_results}

\begin{tabular}{lcccc}

\toprule
                                               &     $\mathrm{ANN}_\mathrm{ref}$          &      $\mathrm{SNN}_\mathrm{ref}$         & $\mathrm{SNN}_{72}$        & $\mathrm{SNN}_{56}$\\
\midrule

\ac{mac} Operations                            &  \num{3600}   &  \num{329600} &  \num{90720}                        &  \num{61600} \\
Latency (\si{\nano \second})                   &  \num{1949}   &  \num{1657}   &  \num{1957}                         &  \num{1657}  \\
Throughput (\si{\kilo Bd})                     &  \num{513}    &  \num{603.5}  &  \num{510.99}                       &  \num{603.5} \\
LUT (\si{\percent})                            &  \num{0.9}    &  \num{45.15}  &  \num{15.9}                         &  \num{12.32} \\
LUT RAM (\si{\percent})                        &  \num{3.27}   &  \num{1.49}   &  \num{1.53}                         &  \num{1.3}   \\
FF (\si{\percent})                             &  \num{1.76}   &  \num{7.55}   &  \num{4.14}                         &  \num{3.37}  \\
BRAM (\si{\percent})                           &  \num{0.8}    &  \num{0.8}    &  \num{0.8}                          &  \num{0.8}   \\
DSP (\si{\percent})                            &  \num{0}      &  \num{0}      &  \num{0}                            &  \num{0}     \\
Total Power (\si{\watt})                    &  \num{1.5}    &  \num{4.3}    &  \num{1.75}                         &  \num{1.6}   \\
Dynamic Power  (\si{\watt})                    &  \num{0.25}   &  \num{2.8}    &  \num{0.25}                         &  \num{0.3}   \\
Dynamic Energy (\si{\micro \joule})            &  \num{0.487}  &  \num{4.64}   &  \num{0.489}                        &  \num{0.497} \\
Energy                                         &               &               &                                     &              \\
Efficiency (\si{\pico \joule \per MAC})        &  \num{135}    &  \num{14.1}   &  \num{5.39}                         &  \num{8.07}  \\
\bottomrule
\end{tabular}
\end{table}

\section{Limitations and future work}
Although our \ac{snn} achieved better communications performance and better energy efficiency per \ac{mac} operation compared to the reference \ac{ann}, it requires more \ac{mac} operations. This highlights the need for advanced learning algorithms for \acp{snn} that reduce their computational complexity while maintaining the same performance. Additionally, despite the custom and flexible \ac{fpga} architecture of the \ac{lif} neuron, the synchronous nature of \acp{fpga} limits its ability to minimize energy consumption, particularly because \acp{snn} are asynchronous and event-based. Future designs should consider further improvements in implementation efficiency. For future work, we plan on replacing the simulated channel with real-world data of different communication scenarios to further validate the robustness of our approach. Furthermore, we aim to investigate other hardware-optimized encoding approaches to reduce the spiking rate.
\section{Conclusion}
In this work, we performed a design space exploration of \ac{snn} based \ac{dfe} to minimize its computational complexity and improve its communication performance. Our findings indicate that \acp{snn} outperformed \acp{ann} in terms of energy efficiency, even on a synchronous and digital platform like \ac{fpga} by a factor of $25\times$ while reducing the \ac{ber} by \SI{52.03}{\percent} on average for an optical communication channel at the same throughput.

\section*{Acknowledgment}

We sincerely thank Prof. Laurent Schmalen and Alexander von Bank from the Communications Engineering Lab (CEL) of Karlsruhe Institute of Technology (KIT) for providing the source code of the \ac{snn}-based \ac{dfe} and for the insightful discussions. 


\bibliographystyle{IEEEtran}
\balance
\bibliography{main}
\vspace{12pt}


\end{document}